*Review Article*

# Chaperones as integrators of cellular networks: changes of cellular integrity in stress and diseases

Robin Palotai[a], Máté S. Szalay[a] and Péter Csermely

*Department of Medical Chemistry, Semmelweis University, Puskin str. 9, H-1088 Budapest, Hungary*

*Summary*

The complex integrity of the cells and its sudden, but often predictable changes can be described and understood by the topology and dynamism of cellular networks. All these networks undergo both local and global rearrangements during stress and the development of diseases. Here we illustrate this showing the stress-induced structural rearrangement of the yeast protein-protein interaction network (interactome). In un-stressed state the yeast interactome is highly compact, and the centrally organized modules have a large overlap. During stress several original modules became more separated, and a number of novel modules also appear. A few basic functions such as the proteasome preserve their central position, however, several functions with high energy demand, such the cell-cycle regulation loose their original centrality during stress. A number of key stress-dependent protein complexes, such as the disaggregation-specific chaperone, Hsp104, gain centrality in the stressed yeast interactome. Molecular chaperones, heat shock, or stress proteins became established as key elements in our molecular understanding of the cellular stress response. Chaperones form complex interaction networks (the chaperome) with each other and their partners. Here we show that the human chaperome recovers the segregation of protein synthesis-coupled and stress-related chaperones observed in yeast recently. Examination of yeast and human interactomes shows that (1) chaperones are inter-modular integrators of protein-protein interaction networks, which (2) often bridge hubs and (3) are favorite candidates for extensive phosphorylation. Moreover, chaperones (4) become more central in the organization of the isolated modules of the stressed yeast protein-protein interaction network, which highlights their importance in the de-coupling and re-coupling of network modules during and after stress. Chaperone-mediated evolvability of cellular networks may play a key role in cellular adaptation during stress and various polygenic and chronic diseases, such as cancer, diabetes or neurodegeneration.

**Running title**  Chaperone networks

[a]Robin Palotai (palotai@linkgroup.hu) and Máté S. Szalay (szalay@linkgroup.hu) are undergraduates of the Budapest University of Technology and Economics, and started their research as members of the Hungarian Research Student Association (www.kutdiak.hu), which provides research opportunities for talented high school students since 1996.

Address correspondence to: Péter Csermely, Department of Medical Chemistry, Semmelweis University School of Medicine, Budapest, P.O. Box 260. H-1444 Hungary. Tel: +361-266-2755 extn.: 4102. Fax: +361-266-6550. E-mail: csermely@puskin.sote.hu

# CELLULAR NETWORKS: TOPOLOGY AND DYNAMISM

*Structural properties of cellular networks*

The network approach dissects the cellular complexity to various elements (also called as vertices, see Table 1), such as proteins, cytoskeletal filaments, cellular organelles, signaling components or enzyme reactions and tries to catalogue those interactions of these elements, which have a relatively high affinity, and, therefore, are measurable with our current, 'traditional' biochemical or high-throughput methods. The interactions (also called as links or edges) often have weights, which reflect their affinity, propensity or probability, and directions, which become especially important in signaling and metabolic networks (Fig. 1; *1–3*).

Cellular networks often form small worlds, where two elements of the network are separated by only four to five other elements as an average (*4*). This proverbial "six steps of separation" is a key feature to limit the distortion of the information, which becomes unmanageable after six transferring steps (*5, 6*). 'Information' in the cellular context is often a conformational change of the participating proteins or other macromolecules, such as RNA-s. Networks of our cells contain hubs, i.e. elements, which have a large number of neighbors. These networks can be dissected to overlapping groups or communities (which we will call modules in this review; *7–10*). Both hubs and network modules are efficient in screening and filtering of the extensive information a cell receives and generates in each second. Hubs can transmit only a minority of the continuously bombarding pieces of information at a given time. Network modules, by definition, have denser intra-modular connections than inter-modular contacts, therefore, keep the incoming information 'trapped' inside the module, and allow its preferential passage to the next module only in special cases, when the network has already been trained to provide a fast transmission of that particular change by previous experience. In this sense all our cells have a unique and special 'personal history', which developed their network configuration to its current status (Fig. 1; *6*).

Cellular networks and their modules often form a hierarchical structure. These 'chain-of-command-type' or pyramid-type organizations are very useful to allow a fast and efficient integration of signaling steps (e.g. in transcription factor networks; *11*), and are helpful in the compartmentalization of cellular metabolism (*12, 13*). Hierarchical networks often display self-similarity, which makes them fractal-like objects (*14*). Cellular networks may possess a 'rich-club', where the hierarchy is configured *via* preferential direct interactions between hubs, or just oppositely, may be 'hub-repulsive', where hubs are separated from each other by other elements. The former is often characteristic to transportation networks, such as metabolic networks, while the latter is often typical to structural networks, such as protein-protein interaction networks (*15, 16*). To add to the complexity of the phenomenon, protein-protein interaction networks may reveal a rich-club phenomenon at an intermediate, but neither at low nor at high levels of their hierarchy, which may reflect their intermediate state between an information transportation network and a structural network (*17*).

At the end of the above initial view, we must warn the readers, that this field is new. Cellular networks have been uncovered only in the last eight years (*1–3*), which may sometimes lead to over-generalizations led by the joy and excitement of a novel understanding of cellular complexity. Moreover, most of our current methods allow only a sampling of the cellular networks resulting in an average network topology from millions of individual cells with opposing 'personal histories'. Therefore, each statement needs to be validated through critical scrutiny of the datasets, sampling procedures and methods of data analysis at each network examined (*18, 19*). As an example of the controversies, which may arise during our current, initial status of understanding, we cite the current 'stratus-(alto)cumulus debate' (*20–23*), where opinions differ, whether the overall organization of the *Saccharomyces cerevisiae* protein-protein interaction network connects network modules by their central, denser core leading to a compact organization, which resembles to the stratus-type clouds, or yeast network modules are connected via their peripheral layers resulting in a more structured organization resembling the cumulus-type clouds.

*Dynamic changes of cellular networks*

Cellular networks display an extremely high dynamism. Not only their links are often rearranged, but due to the synthesis and proteolysis of cellular proteins, a large amount of network elements is continuously vanishing and re-appearing again. An initial and currently often-quoted example of this scenario is the existence of date-hubs and party-hubs, where the date-hubs form complexes with different subsets of their partners at different times and cellular locations, while party-hubs collect their partners and form complexes with all of them simultaneously (*20, 24–26*). Date hubs – logically – usually have a single binding surface, while party hubs are multi-interface proteins (*24*). Date hubs contain more disordered regions, while party hubs have a larger tendency to form a rich-club (*25*). Network modules of the yeast interactome may be dissected to static and dynamic modules, when gene expression changes are taken into account. The pathway structure of static modules is more redundant, which allows a faster evolution and larger tolerance of gene expression noise. On the contrary, dynamic modules help the condition-dependent, flexible regulation of cellular responses (*26*).

Stress may induce a decrease in the strength and number of links, which leads to a gradual detachment of network modules from each other. The overlap decreases between modules leading to simpler, less regulated, and more specialized cellular functions (*27, 28*). As a related study, Luscombe *et al.* (*29*) examined the topology of yeast transcriptional signaling sub-networks of 142 transcription factors and 3,420 target genes in five different cellular conditions. The stress-response was governed by a simplified sub-network, which had a shorter diameter and was characterized by large hubs, which probably behaved as integrators of the re-programmed cellular response. On the contrary, the cell-cycle was governed by a highly interwoven, complex structure indicating a multistage internal program (*29*).

The above, stress-induced topological differences were largely recovered, when we compared the overlapping modular structure of the yeast protein-protein interaction network using the NodeLand version of our recently developed ModuLand method family (*10* and Kovacs *et al.*, in preparation) and a high-confidence dataset of 2,444 yeast proteins and their 6,271 interactions (*25*), where stress was modeled by the re-adjustment of protein abundance values taking into account the average of gene expression changes in 65 experiments of 13 diverse stress conditions (such as heat shock, oxidative and reductive stress, hypo- and hyper-osmotic stress, diauxic shift, nitrogen and amino acid starvation, etc.) from the data of Gasch *et al.* (*30*) as described in the legend of Fig. 2. The height of the peaks on Fig. 2 represent the centrality of the given yeast protein in the yeast interactome. Under normal growth conditions ('un-stressed state', Fig. 2A), the interactome is highly compact, and the centrally organized modules have a large overlap. On the contrary, in the stressed state (Fig. 2B, which is an interactome reflecting the average changes of protein abundance in 13 specific stressful conditions), modules became separated, and their overlap decreases. In the stressed interactome peaks (meaning local centralities) appear at several novel positions. Consequently, the unstressed interactome contains much less modules than the stressed interactome (42 compared to 117 modules in the current analysis). This reflects the emergence of specific protein complexes providing the adaptation to one or another specific stressful event. Applying the 'stratus/cumulus nomenclature', the un-stressed yeast interactome resembles more to a stratus-type, while the stressed interactome to a cumulus-type organization. This suggestion is strengthened by the fact that the number of hub-hub interactions (where hub is defined as an element having 8 or more weighted interactions after Ref. *25*) is decreased to less than half in the interactome of the stressed yeast cell, if compared to the interactome of the yeast cell under normal growth conditions (182 hub-hub interactions *versus* 494 in the un-stressed yeast interactome). This dynamism may explain the variable results obtained before (*20–23*). Further, more detailed examinations are necessary to prove or disprove this assumption.

The central (peak) protein is identical in several modules of the un-stressed and stressed interactomes. Modules, which preserve their centrality in both un-stressed and stressed conditions, are organized around the proteasome, the nuclear transport complex and actin-regulatory proteins. During stress the regulatory complex of the proteasome becomes rearranged, the cell-cycle and rRNA synthesis become suppressed, while the GCN4 stress-dependent transcription complex, the damaged protein-label, ubiquitin, the chemical chaperone-, trehalose-synthesizing complex, the disaggregation-specific chaperone, Hsp104 and the stress surviving cAMP-kinase pathway gain centrality in the interactome (Fig. 2. and data not shown). All these rearrangements emphasize the energy-sparing behavior and point towards the activation and emerging centrality of a multifaceted protection-machinery of the yeast cell during stress. However, we must warn that the illustrative analysis above took into account approximately the strongest 6% of known, and possibly 3% of total yeast protein-protein interactions. The comparison of un-stressed and stressed states probably cancels some of the errors caused by this biased data-set, however, the results certainly do not reflect the whole interaction-complexity. Additionally, the illustrative analysis gave an averaged picture of 13 divergent stress conditions, where individual states may vary. A more detailed analysis (Csermely and co-workers, in preparation) will certainly reveal more details than the initial attempt presented in this review to illustrate the dynamism of modular organization during stress.

The above 'simplification/specialization' duo of the yeast interactome during stress resembles to an accelerated and reversible version of the reductive evolution of symbiotic organisms. In this latter scenario the engulfment by the host provides a safe and stable environment for the 'guest', e.g. a parasite (*31*). In both processes major segments of the original networks become attenuated parallel with a specialization of the network structure for a specific set of environmental conditions provided by either the stress or the host. This network simplification gives a more rigid structure, where most of the original universal and flexible adaptation strategies were temporarily or irreversibly lost.

If the cell experiences an increasing amount of stress, its networks may undergo topological phase transitions. Plenty of resources allow a high link-density keeping contact-preference low, and resulting in a random network-type final configuration. During stress, discrimination between network elements, and contact preferences will occur, and increasingly strong hubs will appear. In an extreme case, the network may be switched to a star-network, where the 'winner hub takes all', and an extremely centralized, highly hierarchical structure develops. With a further reduction of the resources the star-network collapses and a number of isolated, small sub-graphs will be formed. This corresponds to the death of the former gross structure. The latter, disintegration-type topological phase transition may be preceded by quarantining the most damaged modules of the network, and might accompany various forms of programmed cell death (*6, 28, 32, 33*). Currently the above scenario awaits support by experimental evidence. However, the appearance of central, star-like hubs in stress-related sub-networks of yeast transcription factor networks (*29*) supports the possible existence of these rearrangements.

## CHAPERONES AS NETWORK INTEGRATORS

Stress provokes the activation and extensive synthesis of molecular chaperones, many of which are also called heat shock proteins, or stress proteins, abbreviated like "Hsp"-s. Hsp-s provide a general response to stress by repairing damaged proteins. Several chaperones are often abbreviated as "Hsc"-s referring to their heat shock cognate protein status. These chaperones are continuously present in the cells, and assist in protein synthesis, the unfolding/refolding steps of protein transport and structural rearrangements of e.g. the nucleus, as well as participate in the triggering of various signaling steps by releasing the respective kinases or other signaling proteins. Chaperones never work alone, but form large complexes with each other and with their co-chaperones (which we call as chaperome after ref. 37; *34–39*).

The currently available human chaperome is shown on Fig. 3. The core domain contains 14 chaperone molecules, while the periphery has 15 chaperone isoforms. Most of Hsp60 chaperones are in the core, while most of the small heat shock proteins are in the periphery. The Hsp70 and Hsp90 chaperones are divided between the two. Analysis of the individual chaperones reveals that the core domain of the human chaperome corresponds to the yeast "CLIPS-chaperones" (chaperones linked to protein synthesis), while the various subgroups of the periphery reflect the functional divergence of "HSP-chaperones" (stress-induced chaperones) as dissected by Albanese *et al.* (*38*). Human CLIPS-like chaperones are mostly cognate proteins, which are either expressed continuously in human cells, or are brain/testis-specific isoforms of their major counterparts. Human HSP-type chaperones are involved in the stress response, and in the re-programming the cells in malignant transformation or viral infection. The recovery of the major division of the yeast chaperone-arrangement in the human chaperome suggests the evolutionary conservation of this functional distinction.

### *Chaperones as inter-modular connections between hubs*

Molecular chaperones bind and release a large variety of damaged proteins. This is made possible by a large promiscuity in the chaperone-client interactions. Consequently, chaperones form low affinity, dynamic temporary interactions, e.g. weak links in cellular networks (*6, 39, 40*). Chaperones have a large number of hubs among their neighbors in the yeast interactome (*40*). The abundance of hubs as chaperone-neighbors gives a central position to molecular chaperones in the protein-protein interaction network, which may help the chaperone-mediated cross-talk between divergent cellular pathways. Most importantly, chaperones are inter-modular elements of both protein-protein interaction and membrane organelle networks, assembling the modular structure of the cell (*28, 40*). This notion is supported by the observation that chaperones are enriched in date-hubs (*26*), pointing towards their integrative role of various cellular functions. The maximal number of potential phosphorylation sites on various chaperones is color-coded on Fig. 3. The multitude of interacting kinases and potential phosphorylation sites is an additional piece of evidence indicating the regulatory role of chaperones in the organization of human protein-protein interaction networks.

*Role of chaperones in stress-induced network rearrangements*

During stress, chaperones become increasingly occupied by damaged proteins causing a so-called 'chaperone overload' (*41*). Chaperone inhibition may lead to a de-coupling of network modules both in protein-protein interaction networks and in the mitochondrial-ER organelle network. These provide additional safety measures for the cell, since de-coupling of modules may stop the propagation of network damage at the modular boundaries (*6, 28, 42, 43*). Chaperones are marked with red color in Fig. 2. In average molecular chaperones had an approximately 20% higher position in the stressed yeast interactome (Fig. 2B) than in the un-stressed network (Fig. 2A) indicating an increased chaperone-centrality during stress. In other words, chaperones may gain an even more important role in the connection of various segments of the interactome during stress, which suggests a novel sense of the regulatory role of the 'chaperone-overload' i.e. the balance between damaged proteins and available chaperones (*41*). The most central protein of the stressed interactome on Fig. 2B is the damaged protein disaggregating chaperone, Hsp104. Hsp104 forms a complex with a number of other chaperones and co-chaperones, interacts with the Sup35 yeast prion, associates with the ERAD protein degradation pathway of the endoplasmic reticulum, and interacts with the nuclear pore complex. This set of key partners propels Hsp104 as a connector of two out of the three most central modules (the proteasome and the nuclear complex), which explains its extreme centrality. While, the role of Hsp104 in Sup35 regulation and in ERAD protein degradation is well known (*44, 45*), the demonstration of the functional role of Hsp104 in the protection of nuclear pore complexes during stress awaits experimental evidence.

Importantly, while this review was under preparation, a comprehensive study of yeast Hsp90-networks under normal growth conditions and elevated temperature (*36*) recovered many key features of our former and current, illustrative analyses showing that (1) Hsp90 neighbors contained a higher than expected number of hubs (*39*); (2) Hsp90 complexes were rather labile pointing out a preponderance of low-affinity interactions (*6, 39, 40*), and (3) stressed Hsp90 network was more diverse and structured than that under normal growth conditions (Fig. 2), where it is centered around transport processes (data not shown).

When the stress is over, the low affinity and promiscuity of chaperone interactions may provide an efficient tool for the remodeling of the modular structure of the reassembling cellular networks. This may be an underlying reason of the regulatory role of Hsp90 and other chaperones in the evolvability of complex systems (*6, 39, 46*). The recent data of Bobula *et al.* (*47*), which show that a genome-wide mutagenesis with a synthetically harmful mutation screen with the yeast Hsp40/Hsp70 chaperone complex did not recover the usual substrates of this chaperone-machinery, also point towards a network-type explanation of evolvability-regulation.

## IMPLICATIONS IN THE THERAPY OF CHRONIC DISEASES: CANCER, DIABETES, NEURODEGENERATION

Repeated stress induces an accumulating damage in cellular networks and parallel with this 'wears out' the cellular adaptation capacity. This exceedingly happens in chronic disease and during aging (*6, 41*). An additional type of danger is raised by the fact that disease and ageing induce a generally higher noise level (*48*). If disease and age-induced noise is accompanied by the extra, stress-generated noise, it may well go beyond the tolerable threshold, and may induce an 'error-catastrophe'.

Efficient repair of the multiple rearrangements and defects of disease-, ageing- and stress-affected cellular networks is better provided by multi-target drugs than by the 'magic bullets' of traditional drug design. Moreover, the low-affinity binding of multi-target drugs eases the constraints of druggability, and significantly increases the size of the druggable proteome. These effects tremendously expand the number of potential drug targets, and will introduce novel classes of multi-target drugs with smaller side effects and toxicity. In fact, many herbal teas, traditional medicines, nutrients, micro-nutrients, vitamins, and phytonutrients act as multi-target compounds interacting with various cellular networks with a low affinity (*49, 50*).

Due to the multitude of chaperone-mediated interactions and their central role in network integration, chaperone-modulators are excellent, *bona fide* examples of multi-target drugs. Indeed, chaperone substitution (in the form of chemical chaperones, *51*), the help of chaperone induction (*52*) and chaperone inhibition (*53*) are all promising therapeutic strategies involving an increasing number of multi-target drugs acting on the chaperome (*54*).

## SUMMARY AND PERSPECTIVES

The plethora of information in various data-sets and our emerging knowledge on network topology and dynamics provide a unique chance to understand the rearrangements of cellular networks during stress and disease. It is especially intriguing to assess the long-thought integrative role of molecular chaperones at the level of the whole cell, and organism. The highlights of our current knowledge can be summarized as follows:

- In un-stressed state the yeast interactome is highly compact, and the centrally organized modules have a large overlap. During stress the original modules became separated, and novel modules appear.
- Examination of yeast and human interactomes shows that (1) chaperones are inter-modular integrators of protein-protein interaction networks, which (2) often bridge hubs and (3) are favorite candidates for extensive phosphorylation. Moreover, chaperones (4) become more central in the organization of the isolated modules of the stressed yeast protein-protein interaction network, which highlights their importance in the de-coupling and re-coupling of network modules during and after stress. The human chaperome recovers the segregation of protein synthesis-coupled and stress-related chaperones observed in yeast recently.

However, a number of key issues have not been tackled yet both from the theoretical and the experimental points of view.

- We are at the very beginning to understand stress-induced network rearrangements: the exploration of various stress-conditions, as well as parallel data-sets showing the differences between protein-protein interaction, organelle and functional cellular networks are missing. The exploration of topological phase transitions of cellular networks by comparing their topology in extremely resource-rich and resource-poor environments awaits experimentation. Our knowledge on the re-establishment, re-building of cellular networks after stress is practically zero.
- The comparison of chaperomes of various organisms will reveal a lot of exciting and heretofore uncovered functions of these key proteins. Even more importantly, we will have a novel view on their integrative functions, which is a typical emergent network phenomenon, which can not be guessed from our current, fragmented knowledge but needs a global picture of the whole interactomes and other cellular networks.
- We need a much better understanding of cellular network changes in disease and ageing. Besides chaperones-related therapies the design of efficient therapeutic interventions to help cellular networks to cope with stress is missing.

We are quite certain that the re-arrangements of stressed networks and the emergent properties of chaperomes will give a lot of excitement and pleasure in the near future. As a result of these studies the emergence of network-based therapies is expected, where entirely novel target-sets of multi-target drugs will be identified using our knowledge on the vulnerable points (hot-spots) of cellular networks in stress, disease and ageing.

## ACKNOWLEDGEMENTS

The authors would like to thank Huba J. Kiss (University of Pécs) for collecting the human chaperome data, the anonymous referee and members of the LINK-group (www.linkgroup.hu) for helpful suggestions. Work in the authors' laboratory was supported by research grants from the Hungarian National Science Foundation (OTKA-K69105), from the EU (FP6-016003) and by the Hungarian National Research Initiative (NKFP-1A/056/2004 and KKK-0015/3.0).

## REFERENCES

1. Barabasi, A. L., and Oltvai, Z. N. (2004) Network biology: understanding the cell's functional organization. *Nat. Rev. Genet.* **5**, 101–113.
2. Zhu, X., Gerstein, M., and Snyder, M. (2007) Getting connected: analysis and principles of biological networks. *Genes Dev.* **21**, 1010–1024.
3. Almaas, E. (2007) Biological impacts and context of network theory. *J. Exp. Biol.* **210**, 1548–1558.
4. Watts, D. J., and Strogatz, S. H. (1998) Collective dynamics of 'small-world' networks. *Nature* **393**, 440–442.

5. Maslov, S., and Ispolatov, I. (2007) Propagation of large concentration changes in reversible protein-binding networks. *Proc. Natl. Acad. Sci. USA* **104**, 13655–13660.
6. Csermely, P. (2006) *Weak links: a universal key for network diversity and stability*. Springer Verlag, Heidelberg.
7. Barabasi, A. L., and Albert, R. (1999) Emergence of scaling in random networks. *Science* **286**, 509–512.
8. Albert, R. (2005) Scale-free graphs in cell biology. *J. Cell Sci.* **118**, 4947–4957.
9. Palla, G., Derenyi, I., Farkas, T., and Vicsek, T. (2005) Uncovering the overlapping community structure of complex networks in nature and society. *Nature* **435**, 814–818.
10. Kovács, I. A., Szalay, M. S., Csermely, P., and Korcsmáros, T. (2006) Method for analyzing the fine structure of networks. Patent application number: PCT/IB2007/05047.
11. Yu, H., and Gerstein, M. (2006) Genomic analysis of the hierarchical structure of regulatory networks. *Proc. Natl. Acad. Sci. USA* **103**, 14727–14731.
12. Ravasz, E., Somera, A. L., Mongru, D. A., Oltvai, Z. N., and Barabasi, A.-L. (2002) Hierarchical organization of modularity in metabolic networks. *Science* **297**, 1551–1555.
13. Sales-Pardo, M., Guimerá, R., Moreira, A. A., and Amaral, L. A. N. (2007) Extracting the hierarchical organization of complex systems. *Proc. Natl. Acad. Sci. USA* **104**, 15224–15229.
14. Song, C., Havlin, S., and Makse, H. A. (2005) Self-similarity of complex networks. *Nature* **433**, 392–395.
15. Colizza, V., Flammini, A., Serrano, M. A., and Vespignani, A. (2006) Detecting rich-club ordering in complex networks. *Nat. Phys.* **2**, 110–115.
16. Guimerá, R., Sales-Pardo, M., and Amaral, L. A. N. (2007) Classes of complex networks defined by role-to-role connectivity profiles. *Nat. Phys.* **3**, 63–69.
17. McAuley, J. J., da Fontoura Costa, L., and Caetano, T. S. (2007) The rich-club phenomenon across complex network hierarchies. *Appl. Phys. Lett.* **91**, 084103.
18. Arita, M. (2004) The metabolic world of *Escherichia coli* is not small. *Proc. Natl. Acad. Sci. USA* **101**, 1543–1547.
19. Tanaka, R., Yi, T. M., and Doyle, J. (2005) Some protein interaction data do not exhibit power law statistics. *FEBS Lett.* **579**, 5140–5144.
20. Han, J. D., Bertin, N., Hao, T., Goldberg, D. S., Berriz, G. F., Zhang, L. V., Dupuy, D., Walhout, A. J., Cusick, M. E., Roth, F. P., and Vidal, M. (2004) Evidence for dynamically organized modularity in the yeast protein-protein interaction network. *Nature* **430**, 88–93.
21. Batada, N. N., Reguly, T., Breitkreutz, A., Boucher, L., Breitkreuz, B. J., Hurst, L. D., and Tyers, M. (2006) Stratus not altocumulus: A new view on the yeast protein-protein interaction network. *PLoS Biol.* **4**, e317.
22. Bertin, N., Simonis, N., Dupuy, D., Cusick, M. E., Han, J. D., Fraser, H. B., Roth, F. P., and Vidal, M. (2007) Confirmation of organized modularity in the yeast interactome. *PLoS Biol.* **5**, e153.
23. Batada, N. N., Reguly, T., Breitkreutz, A., Boucher, L., Breitkreuz, B. J., Hurst, L. D., and Tyers, M. (2006) Still stratus not altocumulus: further evidence against the date/party hub distinction. *PLoS Biol.* **5**, e154.
24. Kim, P. M., Lu, L. J., Xia, Y., and Gerstein, M. B. (2006) Relating three-dimensional structures to protein networks provides evolutionary insights. *Science* **314**, 1938–1941.
25. Ekman, D., Light, S., Björklund, A. K., and Elofsson, A. (2006) What properties characterize the hub proteins of the protein-protein interaction network of *Saccharomyces cerevisiae*? *Genome Biol.* **7**, R45.
26. Komurov, K., and White, M. (2007) Revealing static and dynamic modular architecture of the eukaryotic protein interaction network. *Mol. Systems Biol.* **3**, 110.
27. Szalay, M. S., Kovács, I. A., Korcsmáros, T., Böde. C., and Csermely, P. (2007) Stress-induced rearrangements of cellular networks: consequences for protection and drug design. *FEBS Lett.* **581**, 3675–3680.
28. Soti, C., Pal, C., Papp, B., and Csermely, P. (2005) Chaperones as regulatory elements of cellular networks. *Curr. Op. Cell Biol.* **17**, 210–215.
29. Luscombe, N. M., Babu, M. M., Yu, H., Snyder, M., Teichmann, S. A., and Gerstein, M. (2004) Genomic analysis of regulatory network dynamics reveals large topological changes. *Nature* **431**, 308–312.
30. Gasch, A. P., Spellman, P. T., Kao, C. M., Carmel-Harel, O., Eisen, M. B., Storz, G., Botstein, D., and Brown, P. O. (2000) Genomic expression programs in the response of yeast cells to environmental changes. *Mol. Biol. Cell* **11**, 4241–4257.
31. Pál, C., Papp, B., Lercher, M. J., Csermely, P., Oliver, S. G., and Hurst, L. D. (2006) Chance and necessity in the evolution of minimal metabolic networks. *Nature* **440**, 667–670.
32. Derenyi, I., Farkas, I., Palla, G., and Vicsek, T. (2004) Topological phase transitions of random networks. *Physica A* **334**, 583–590.
33. Soti, C., Sreedhar, A. S., and Csermely, P. (2003) Apoptosis, necrosis and cellular senescence: chaperone occupancy as a potential switch. *Ageing Cell* **2**, 39–45.
34. Macario, A. J., and Conway de Macario, E. (2005) Sick chaperones, cellular stress, and disease. *N. Engl. J. Med.* **353**, 1489–1501.
35. Blatch, G, L. (2007) Ed. *Networking of chaperones by co-chaperones*. Springer Verlag, Heidelberg.

36. McClellan, A. J., Xia, Y., Deutschbauer, A. M., Davis, R.W., Gerstein, M., and Frydman, J. (2007) Diverse cellular functions of the Hsp90 molecular chaperone uncovered using systems approaches. *Cell* **131**, 121–135.
37. Wang, X., Venable, J., LaPointe, P., Hutt, D. M., Koulov, A. V., Coppinger, J., Gurkan, C., Kellner, W., Matteson, J., Plutner, H., Riordan, J. R., Kelly, J. W., Yates, J. R. 3rd, and Balch, W. E. (2006) Hsp90 cochaperone Aha1 downregulation rescues misfolding of CFTR in cystic fibrosis. *Cell* **127**, 803–815.
38. Albanese, V., Yam, A. Y., Baughman, J., Parnot, C., and Frydman, J. (2006) Systems analyses reveal two chaperone networks with distinct functions in eukaryotic cells. *Cell* **124**, 75–88.
39. Korcsmaros, T., Kovacs, I. A., Szalay, M. S., and Csermely, P. (2007) Molecular chaperones: the modular evolution of cellular networks. *J. Biosci.* **32**, 441–446.
40. Csermely P. (2004) Strong links are important, but weak links stabilize them. *Trends Biochem. Sci.* **29**, 331–334.
41. Csermely P. (2001) Chaperone-overload as a possible contributor to "civilization diseases": atherosclerosis, cancer, diabetes. *Trends Genet.* **17**, 701–704.
42. Szabadkai, G., Bianchi, K., Varnai, P., De Stefani, D., Wieckowski, M. R., Cavagna, D., Nagy, A. I., Balla, T., and Rizzuto, R. (2006) Chaperone-mediated coupling of endoplasmic reticulum and mitochondrial Ca2+ channels. *J. Cell Biol.* **175**, 901–911.
43. Lewandowska, A., Gierszewska, M., Marszalek, J., and Liberek. K. (2006) Hsp78 chaperone functions in restoration of mitochondrial network following heat stress. *Biochim. Biophys. Acta* **1763**, 141–151.
44. Shorter, J., and Lindquist, S. (2004) Hsp104 catalyzes formation and elimination of self-replicating Sup35 prion conformers. *Science* **304**, 1793–1797.
45. Taxis, C., Hitt, R., Park, S. H., Deak, P. M., Kostova, Z., and Wolf, D. H. (2003) Use of modular substrates demonstrates mechanistic diversity and reveals differences in chaperone requirement of ERAD. *J. Biol. Chem.* **278**, 35903–35913.
46. Rutherford, S. L., and Lindquist, S. (1998) Hsp90 as a capacitor for morphological evolution. *Nature* **396**, 336–342.
47. Bobula, J., Tomala, K., Jez, E., Wloch, D. M., Borts, R. H., and Korona, R. (2006) Why molecular chaperones buffer mutational damage: a case study with a yeast Hsp40/70 system. *Genetics* **174**, 937–944.
48. Goldberger, A. L., Amaral, L. A. N., Hausdorf, J. M., Ivanov, P. C., Peng, C.-K., and Stanley, H. E. (2002) Fractal dynamics in physiology: alterations with disease and ageing. *Proc. Natl. Acad. Sci. USA* **99**, 2466–2472.
49. Csermely, P., Agoston, V., and Pongor, S. (2005) The efficiency of multi-target drugs: the network approach might help drug design. *Trends Pharmacol. Sci.* **26**, 178–182.
50. Korcsmáros, T., Szalay, M. S., Böde. C., Kovács, I. A., and Csermely, P. (2007) How to design multi-target drugs: Target-search options in cellular networks. *Expert Op. Drug Discov.* **2**, 1–10.
51. Papp, E., and Csermely, P. (2006) Chemical chaperones: mechanisms of action and potential use. *Handb. Exp. Pharmacol.* **172**, 405–416.
52. Vigh, L., Literati, P. N., Horvath, I., Torok, Z., Balogh, G., Glatz, A., Kovacs, E., Boros, I., Ferdinandy, P., Farkas, B., Jaszlits, L., Jednakovits, A., Koranyi, L., and Maresca, B. (1997) Bimoclomol: a nontoxic, hydroxylamine derivative with stress protein-inducing activity and cytoprotective effects. *Nat. Med.* **3**, 1150–1154.
53. Xu, W., and Neckers, L. (2007) Targeting the molecular chaperone heat shock protein 90 provides a multifaceted effect on diverse cell signaling pathways of cancer cells. *Clin. Cancer Res.* **13**, 1625–1629.
54. Soti, C., Nagy, E., Giricz, Z., Vigh, L., Csermely, P., and Ferdinandy, P. (2005) Heat shock proteins as emerging therapeutic targets. *Br. J. Pharmacol.* **146**, 769–780.
55. Batagelj, V., and Mrvar, A. (1998) Pajek - Program for Large Network Analysis. *Connections* **21**, 47–57.
56. Ng, A., Bursteinas, B., Gao, Q., Mollison, E., and Zvelebil, M. (2006) pSTIING: a 'systems' approach towards integrating signalling pathways, interaction and transcriptional regulatory networks in inflammation and cancer. *Nucleic Acids Res.* **34**, D527–D534.
57. Linding, R., Jensen, L. J., Ostheimer, G. J., van Vugt, M. A., Jorgensen, C., Miron, I. M., Diella, F., Colwill, K., Taylor, L., Elder, K., Metalnikov, P., Nguyen, V., Pasculescu, A., Jin, J., Park, J. G., Samson, L. D., Woodgett, J. R., Russell, R. B., Bork, P., Yaffe, M. B., and Pawson, T. (2007) Systematic discovery of in vivo phosphorylation networks. *Cell* **129**, 1415–1426.

**Table 1.** A glossary of network-specific expressions

| Expression | Short explanation |
|---|---|
| centrality | Centrality of a network *element* or interaction defines relative importance of the *element* or interaction within the network (for example, how important a person is within a social network or a protein in a cellular network). There are various measures of centrality in network analysis giving numerical figures to characterize this importance from the local structure of the network interactions, global properties of the whole network, or both. In our current paper we refer to the latter, complex understanding of centrality, which takes into account all levels of network structure. |
| chaperome | The protein-protein interaction network of molecular chaperones with each other and with their targets. The chaperome is a segment of the *interactome*, which contains all protein-protein interactions in the respective cell type. |
| element (node, vertex or vertices) | The element is a single building block of a network. The element is also called a vertex in graph theory, site in physics or actor in sociology. Most of the times the element itself is a complex network again, like the elements of cellular protein-protein interactions networks, the individual protein molecules can be perceived as networks of their constituting amino acids or atoms. |
| fractal | Fractal objects are generated by recursive process, where self-similar objects of different size are repeated and repeated again. In nature we are often talking about fractal-like behavior, where the extent of self-similarity is not complete as in pure (and many times extremely beautiful) mathematical fractals. |
| hierarchical network | A hierarchical organization arises in a network, when an *element* has a 'parent' and this 'parent' also has a 'grandparent', like in a family tree. Networks may contain more subtle hierarchies, where network *modules* form a network, where again, network *modules* can be defined, which also form a network, etc. |
| hub | A hub is a highly connected *element* of the network. Usually a hub has more than 1% of total interactions. |
| hub-repulsive | A network is called as hub-repulsive, when its *hubs* are connected to each other with a smaller probability than in a network, which contains the same number of *hubs*, but they are connected randomly. The opposite of a hub-repulsive network contains a *rich-club*. |
| interactome | The protein-protein interaction network of a respective cell type. Please note that the interactome always contains data which were averaged from a high number of individual experiments involving myriads of cells. We do not have yet the means to uncover the interactome of a specific, single cell. |
| modules (network communities, groups) | Modules are groups of network *elements* that are relatively isolated from the rest of the network, and where the *elements* inside the module are functionally and/or physically linked to each other. |
| rich-club | A network has a rich-club, when its *hubs* are connected to each other with a higher probability than in a network, which contains the same number of *hubs*, but they are connected randomly. The opposite of a rich club-containing network is a *hub-repulsive* network. |

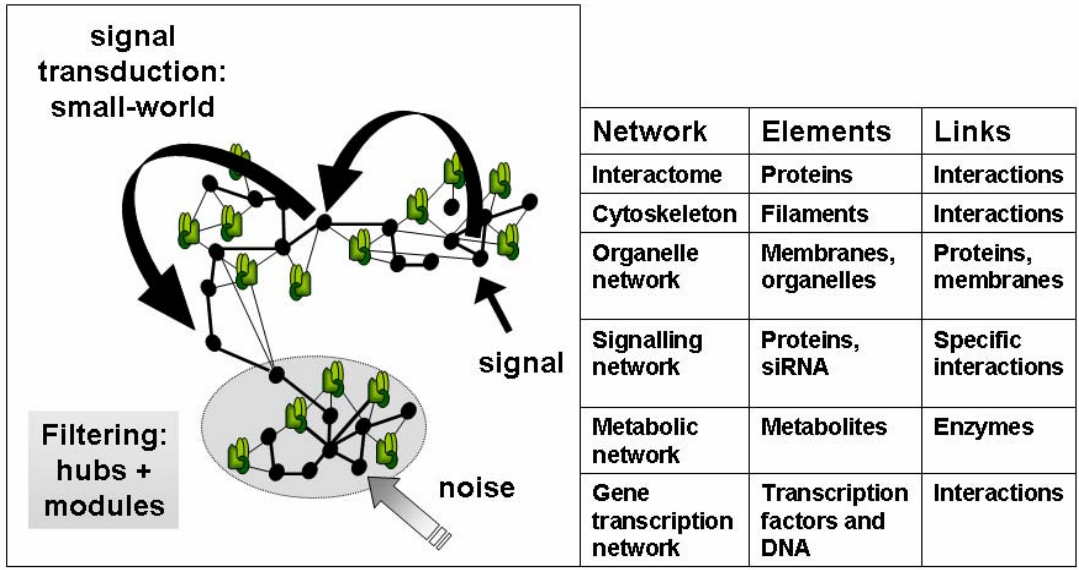

**Figure 1.** Cellular networks and their major functions: information propagation by small-worldness and filtering by hubs and network modules.

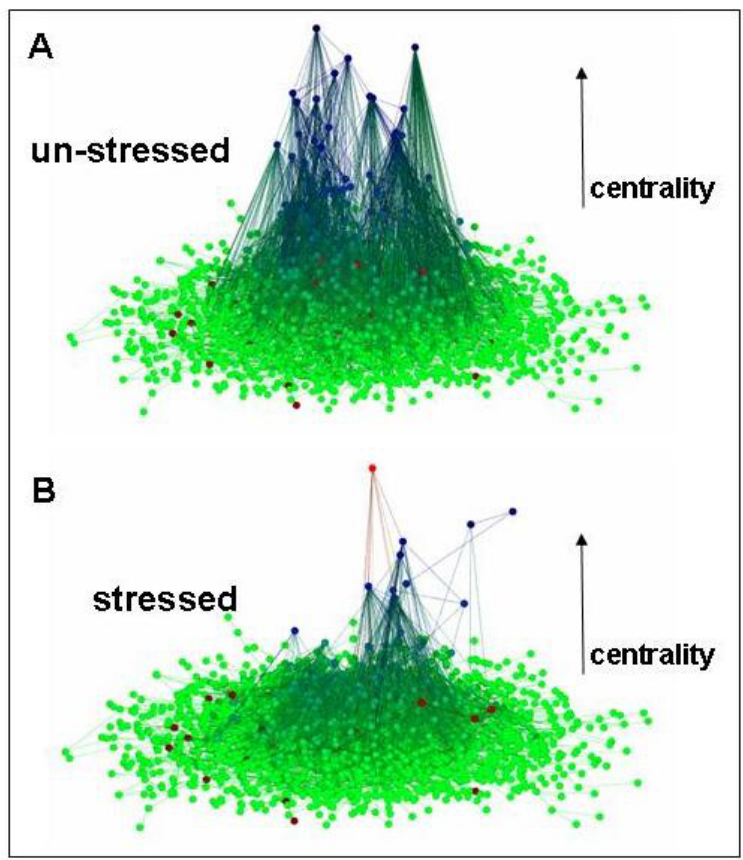

**Figure 2.** Rearrangements of the yeast interactome, and changes of yeast chaperone positions in stress. The figure shows the yeast protein-protein interaction network using the high-confidence dataset of 2,640 yeast proteins and their 6,600 interactions (*25*). The peaks represent network modules detected by the ModuLand method (*10*), and the vertical position corresponds to the centrality of the given protein. The color-scale of light green to dark blue represents increasing centralities, while molecular chaperones were coded by the red color. Colored lines represent the interactions (links) between proteins. Line colors were set according to the color of the end-points. The network was visualized with the modification of the Pajek program (*55*). (A) The un-stressed yeast interactome. (B) The stressed yeast interactome. Stress was modeled by the re-adjustment of the uniform link-weight of 1.0 of the un-stressed data-set taking into account the average of gene expression changes in 65 experiments of 13 diverse stress conditions from the data of Gasch *et al.* (heat shock 5, 15, 30, 40 and 80 min; 37→25°C 15, 30, 45, 60 and 90 min; 0.32 mM hydrogen peroxide 10, 30, 50, 80 and 120 min; 1 mM menadione 10, 30, 50, 105 and 160 min; 2.5 mM dithio-threitol 15, 30, 60, 120 and 480 min; 1.5 mM diamide 5, 20, 30, 50 and 90 min; 1 M sorbitol 5, 15, 45, 60 and 120 min; hypo-osmotic shock 5, 15, 30, 45 and 60 min; amino-acid starvation 0.5, 1, 2, 4 and 6 hours; nitrogen depletion 0.5, 2, 8, 24 and 72 hours; diauxic shift 9.5, 11.5, 13.5, 15.5 and 20.5 hours; YPD growth 2, 6, 10, 24 and 72 hours; YPD stationary phase 0.33, 1, 3, 7 and 22 days; as described in detail in Ref. *30*). The signed sum of minimum two-fold changes in gene expression of the 65 experiments was counted. Weights of respective proteins were assigned as 0.25 or 0.5, if the minimum two-fold decrease was observed between 29 and 22, or between 21 and 12 experiments, respectively. Weights were set as 2 or 4, if the minimum two-fold increase was observed between 10 and 20, or between 21 and 40 experiments, respectively. Link-weights were calculated as the products of their two endpoint weights.

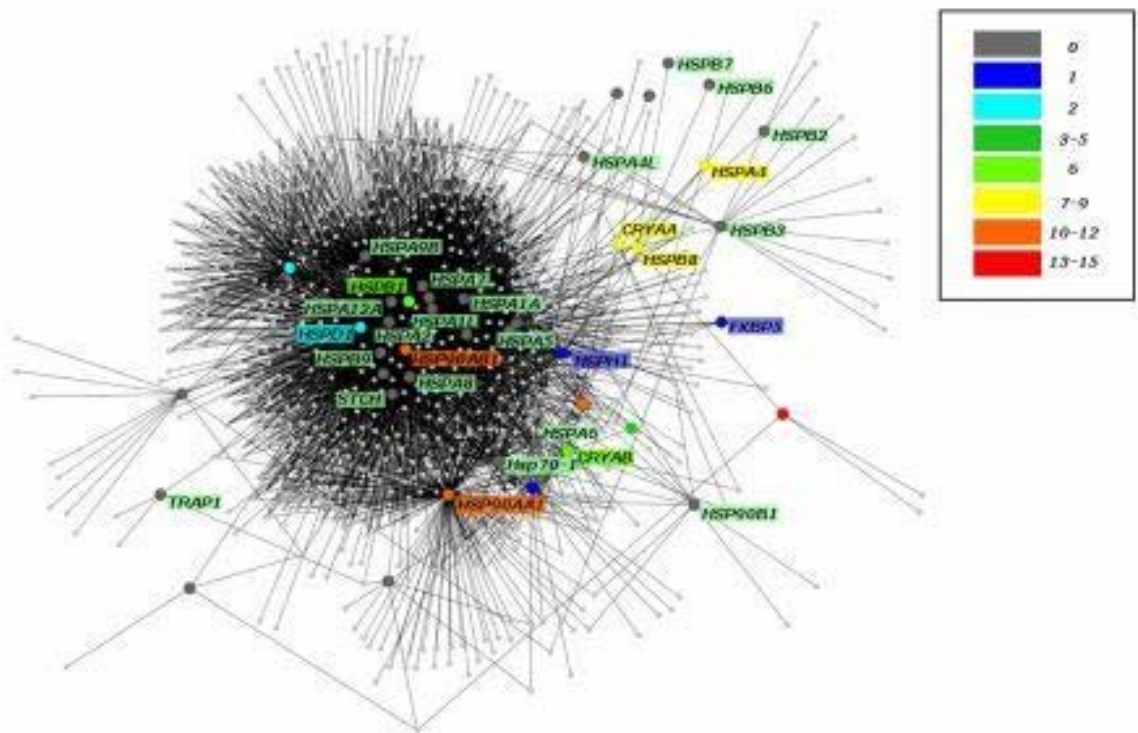

**Figure 3.** The phosphorylated human chaperome. Interacting neighbors of human molecular chaperones were identified from the 16.09.07. release of the pSTIING database (*56,* pstiing.licr.org). Potential phosphorylation sites were obtained from the same release of the NetworKIN database (*57,* networkin.info). Interacting proteins are marked as white small circles, co-chaperones as large grey circles, while chaperones are color-coded according to the number of their potential phosphorylation sites shown in the inset. Grey lines represent protein-protein interactions. The abbreviated names of human molecular chaperones are given with the respective phosphorylation-specific color. The network was visualized with the Pajek program (*55*).